\documentclass[conference]{IEEEtran}
\IEEEoverridecommandlockouts

\usepackage{cite}
\usepackage{amsmath,amssymb,amsfonts}
\usepackage{algorithmic}
\usepackage{graphicx}
\usepackage{textcomp}
\usepackage{xcolor}
\usepackage{booktabs}
\usepackage{multirow}
\usepackage{float}
\usepackage{makecell} 
\def\BibTeX{{\rm B\kern-.05em{\sc i\kern-.025em b}\kern-.08em
    T\kern-.1667em\lower.7ex\hbox{E}\kern-.125emX}}
\begin{document}

\title{Quantifying the Unintentional Islanding Risk: A Comparative Study on Active Distribution Network}

\author{\IEEEauthorblockN{Harshit Nayak}
\IEEEauthorblockA{\textit{Intelligent Electrical Power Grids} \\
\textit{Delft University of Technology}\\
Delft, The Netherlands \\ h.n.nayak@tudelft.nl}
\and
\IEEEauthorblockN{Edoardo Daccò}
\IEEEauthorblockA{\textit{Department of Energy} \\
\textit{Politecnico di Milano}\\
Milan, Italy \\
edoardo.dacco@polimi.it}
\and
\IEEEauthorblockN{Jose Luis Rueda Torres}
\IEEEauthorblockA{\textit{Intelligent Electrical Power Grids} \\
\textit{Delft University of Technology}\\
Delft, The Netherlands \\
J.L.RuedaTorres@tudelft.nl}
\and
\IEEEauthorblockN{Davide Falabretti}
\IEEEauthorblockA{\textit{Department of Energy} \\
\textit{Politecnico di Milano}\\
Milan, Italy \\
davide.falabretti@polimi.it}
}

\maketitle

\begin{abstract}
The growing share of Inverter-Based Resources (IBRs) is changing the operation of modern Active Distribution Networks (ADNs), raising concerns for grid stability, protection, and reliability. In this paper, we investigate the dynamic behavior of a Medium-Voltage (MV) portion of an ADN during the transition from grid-connected to islanded operation. The study considers a Grid-Following (GFL) converter and Synchronous Generators (SGs) operated under two distinct regulation modes: a fixed-setpoint, non-regulating condition representative of present-day distribution networks, and an active frequency/voltage-regulating condition representative of SGs equipped with governor and AVR control. Detailed electromagnetic transient (EMT) models of both technologies have been developed in DIgSILENT PowerFactory that quantifies the sensitivity of island persistence to SG regulation mode and inertia. The results show that when DERs operate at a fixed power setpoint, the risk of forming a sustained, undetected electrical island is limited, since the isolated network drifts out of the protection thresholds within tens of seconds. Conversely, enabling frequency and voltage regulation on the SGs is sufficient to sustain an unintentional island indefinitely without triggering conventional protection, regardless of system inertia. \\
\end{abstract}



\section{Introduction}
The global energy landscape is undergoing a major transition driven by the need to address climate change and reduce greenhouse gas emissions. In 2019 the European Union committed to achieving climate neutrality by 2050, as formalized in the European Green Deal \cite{b1}. Distributed Energy Resources (DERs) offer a sustainable alternative, but their variability and limited predictability introduce new challenges for stability, reliability, and protection design in ADNs \cite{b2}.

In this context, the vast majority of DERs are connected to the grid through power electronic converters, referred to as Inverter-Based Resources, and are broadly divided into Grid-Forming (GFM) and Grid-Following (GFL) converters \cite{b3,b4,b5,b6}. Their operational characteristics differ substantially from those of conventional rotating machines: unlike SGs, which can supply short-circuit currents of five to seven times their rated capacity \cite{b7}, the IBRs can contribute little more than their rated current \cite{b8}, weakening system strength and limiting the ability to support voltage profiles after contingencies \cite{b9}. This difference becomes critical when a portion of the network is unintentionally separated from the main system, and it is further compounded by the regulation mode of the rotating machines that remain in service on the islanded portion.

Existing studies in Table \ref{Table1},  have primarily focused on anti-islanding detection and prevention schemes or intentional islanding involving a single generation technology, typically SGs or GFM-based resources. Although a limited number of studies have validated islanding behavior using real distribution networks, and none have systematically quantified the influence of SG operating modes namely fixed-setpoint operation versus active frequency and voltage regulation on the persistence of unintentional islands. Furthermore, such assessments have not been benchmarked against a GFL inverter, despite the widespread deployment of GFL controlled converters in modern active distribution networks. A detailed analysis of what actually happens to a portion of an
ADN during the transition from grid connected to islanded operation, with both a non-regulating and a regulating rotating
generator mix in service, is still largely missing. Motivated by this gap, the present work investigates the dynamic behavior of a MV portion of an ADN during an unintentional islanding event, triggered by the tripping of the circuit breaker at the beginning of an MV line. After the disconnection, the dispersed generation may temporarily keep the network energized, producing complex voltage and frequency transients that arise from the interaction of the different control systems. Particular attention is devoted to identifying the conditions under which the electrical island persists, a situation strongly discouraged by Distribution System Operators (DSOs) because it can lead to asset deterioration, safety hazards, and out-of synchronism automatic reclosing.

The main contributions of this paper are:
\begin{itemize}
\item Using a real network-validated MV benchmark, this work quantifies the impact of SG regulation mode on unintentional island survivability. It shows that fixed-setpoint operation leads to island collapse within 12.5–21 s, whereas active frequency and voltage regulation enables sustained islanded operation, independent of generation technology.
\item Practical, network-validated tripping-time and persistence thresholds for researchers and DSOs dealing with increasing regulation capability in distribution-connected DERs, and a clear identification of grid-forming converters as the critical next case to investigate.
\end{itemize}

All numerical analyses have been carried out on the model of a real MV distribution network located in Italy, developed in DIgSILENT PowerFactory. ElectroMagnetic Transient (EMT) simulations have been adopted, since they allow a detailed representation of switching phenomena, electromagnetic interactions, and control system responses over very short time scales.

The remainder of the paper is organized as follows. Section \ref{section2} presents the developed dynamic models, with particular focus on the GFL control strategy. Section \ref{section3} describes the case study and test matrix, Section \ref{section4} discusses the numerical results, and Section \ref{section5} draws the main conclusions.

\begin{table}[t]
\label{Table1}
\caption{State-of-the-Art Review and Research Positioning}
\label{tab:relwork}
\centering
\footnotesize
\begin{tabular}{@{}p{0.9cm}p{1.7cm}p{1.4cm}p{1.5cm}p{1.5cm}@{}}
\toprule
\textbf{Ref.} & \textbf{Scope} & \textbf{Real network} & \textbf{Multi-tech.\ benchmark} & \textbf{Quantified thresholds} \\
\midrule
\cite{b10} & Prob.\ risk of unintend.\ islanding & No & No & Statistical \\
\cite{b11} & GFM active detection & No & GFM only & No \\
\cite{b12} & Intentional islanding, SG & Yes & SG only & Partial \\
\cite{b13} & Unintend.\ islanding, GFL/GFM (lab-scale) & No (microgrid testbed) & GFL/GFM only & Partial (NDZ, timing) \\
\cite{b14} & Islanding detection, GFM/GFL & No & GFL/GFM only & No \\
\cite{b15} & Unintend.\ islanding, induction motors & Yes (field) & No & No \\
\makecell{\cite{b16},\\\cite{b17},\\\cite{b18}} & GFM black-start & No & GFM only & No \\
\textbf{This work} & Unintend.\ islanding, SG/GFL & \textbf{Yes} & \textbf{Yes (fixed vs.\ regulating SG)} & \textbf{Yes} \\
\bottomrule
\end{tabular}
\end{table}

\section{Dynamic Modeling of Generation Units}
\label{section2}
This Section presents the dynamic models of the SG and GFL units implemented in DIgSILENT PowerFactory to compare the behavior of different DERs under grid-connected and islanding conditions. All models were calibrated against actual DERs deployed in real ADNs, ensuring a meaningful comparison under identical operating scenarios.

\subsection{Synchronous Generator Modeling}
\begin{figure}[htbp]
\centering
\includegraphics[width=\linewidth]{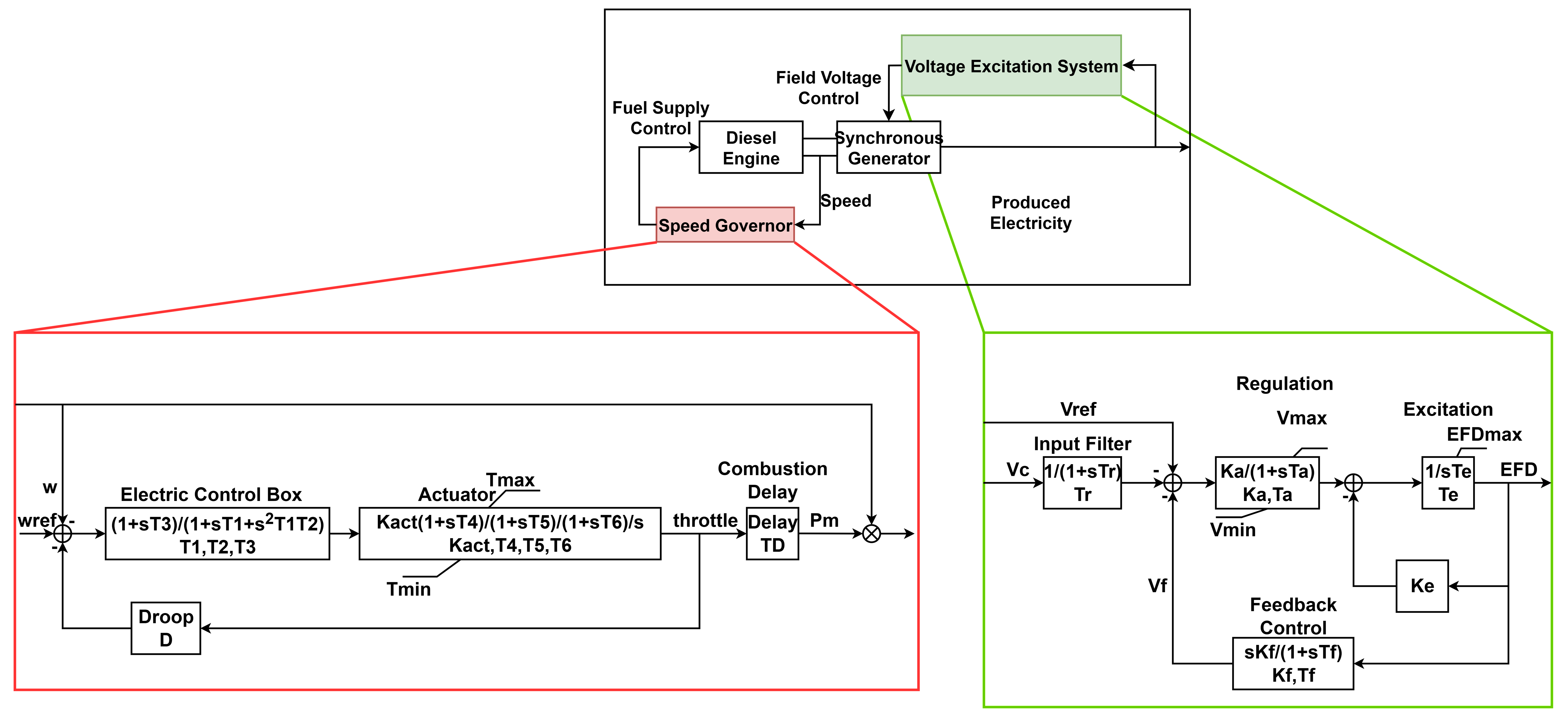}
\caption{Control scheme of implemented synchronous generator.}
\label{fig:SGscheme}
\end{figure}

The implemented SG model replicates the dynamic behavior of a real-world diesel-driven rotating generator. Following standard practice, a three-phase synchronous alternator is adopted, with capability limits defined by stator current, rotor excitation, turbine torque, and synchronism stability margins. A detailed description of the SG's dynamic model and power capabilities is provided in \cite{b12}. The main features of the SG model are Automatic Voltage Regulator (AVR), and speed governor that are summarized below and shown in  Fig-\ref{fig:SGscheme}.
The speed governor follows the classical Woodward diesel controller structure, operating in both isochronous (constant frequency) and droop (frequency-dependent) modes to assess different frequency regulation strategies. It accounts for droop tuning, actuator saturation, and combustion delay, which together set the mechanical torque supplied to the alternator. The AVR uses a static excitation system aligned with IEEE standards, combining voltage regulation via a proportional-integral controller, excitation voltage integration with saturation, and derivative feedback to improve transient response.

\subsection{Grid Following Inverter Modeling}
\begin{figure}[htbp]
\centering
\includegraphics[width=\linewidth]{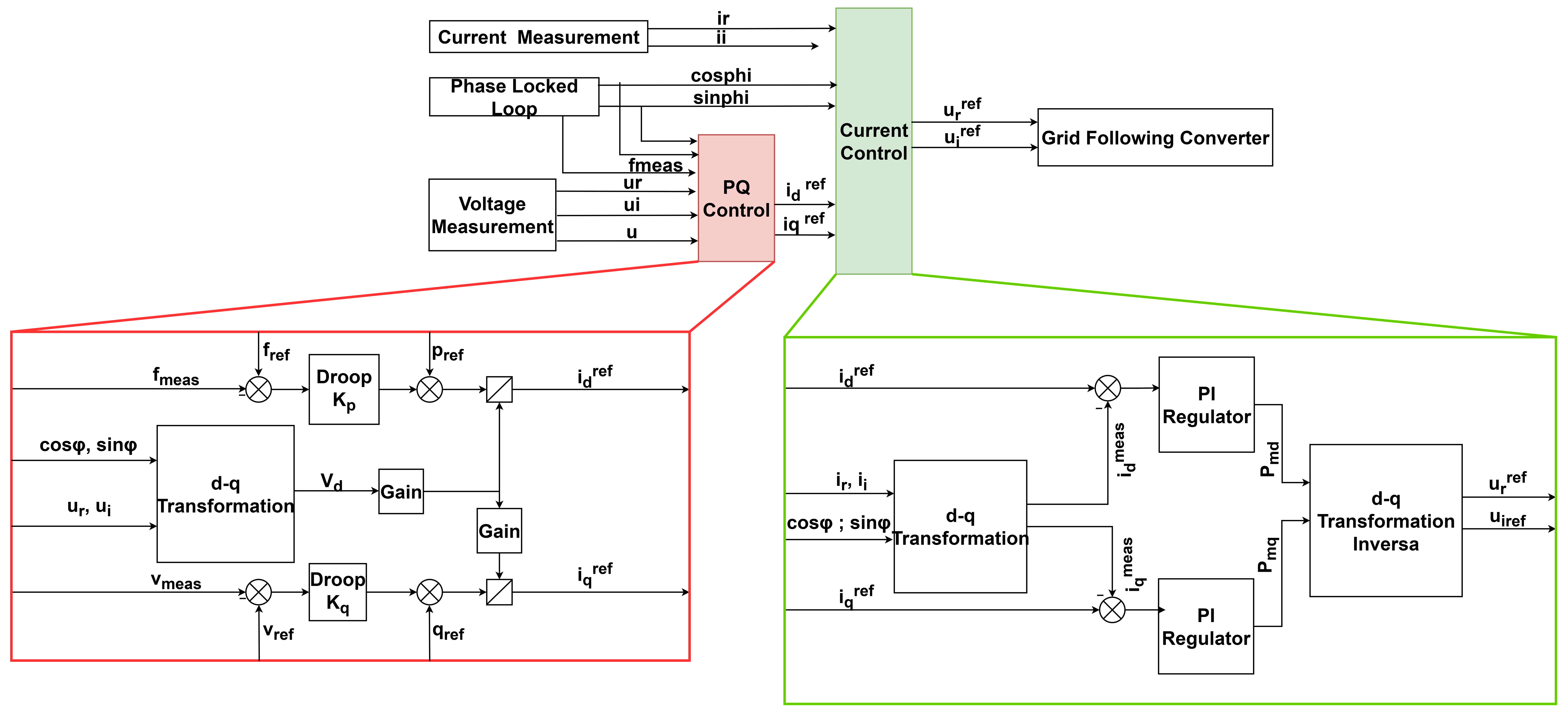}
\caption{The implemented GFL control scheme.}
\label{fig:GFLscheme}
\end{figure}

In line with real-world applications, the developed GFL model includes the  Phase Locked Loop (PLL) together with the PI control strategies governing the current and power control loops \cite{b20,b21}. The GFL control scheme, shown in Fig.~\ref{fig:GFLscheme}, combines built-in DIgSILENT PowerFactory blocks of current, voltage measurements and the PLL with user-defined blocks implementing inner current and outer power control in a cascaded structure. The GFL inverter itself is modeled through the built-in "static generator" module, whose input quantities are set by these custom regulators.

\subsubsection{Outer Power Control}
The power controller regulates the active and reactive current
references from the frequency and voltage measurements provided by
the PLL and voltage measurement blocks. The frequency deviation from
its nominal reference $f_{\text{ref}}$ is
\begin{equation}
    \Delta f = f_{\text{ref}} - f,
    \label{eq:domega}
\end{equation}
and the direct-axis current reference follows an active-power droop
law,
\begin{equation}
    i_d^{ref} = p_{ref} + K_p\,\Delta f,
    \label{eq:id_star}
\end{equation}
where $K_p$ is the active-power droop coefficient. Similarly, the
voltage-magnitude deviation is
\begin{equation}
    \Delta|v| = |v|_{\text{ref}} - |v|,
    \label{eq:dv}
\end{equation}
and the quadrature-axis current reference follows a reactive-power
droop law,
\begin{equation}
    i_q^{ref} = q_{ref} + K_q\,\Delta|v|,
    \label{eq:iq_star}
\end{equation}
with $K_q$ the reactive-power droop coefficient. The real and
imaginary voltage components $v_d$, $v_q$ obtained from the
measurement blocks are used to decouple the active- and
reactive-current channels.
\subsubsection{Inner Current Control}
The converter current measurement $i_r$, $i_i$ is converted into synchronous $dq$ reference frame using the grid voltage angle $\theta$ obtained from the PLL,
\begin{equation}
    i_d = i_{r}\cos\theta + i_{i}\sin\theta, \qquad
    i_q = -i_{r}\sin\theta + i_{i}\cos\theta.
    \label{eq:idq_meas}
\end{equation}
The errors between the reference currents $i_d^{ref}$, $i_q^{ref}$ and
their measured counterparts $i_d^{meas}$, $i_q^{meas}$ are processed by two
independent proportional-integral (PI) regulators,
\begin{equation}
    v_d^{*} = K_{i,c}\!\left(1 + \frac{1}{sT_{i,c}}\right)
        \left(i_d^{ref} - i_d^{meas}\right),
    \label{eq:vd_star}
\end{equation}
\begin{equation}
    v_q^{*} = K_{i,c}\!\left(1 + \frac{1}{sT_{i,c}}\right)
        \left(i_q^{ref} - i_q^{meas}\right),
    \label{eq:vq_star}
\end{equation}
where $K_{i,c}$ and $T_{i,c}$ are the current-controller gain and
integrator time constant, respectively (Table~\ref{tab:der_params}).
The resulting $dq$-frame voltage references $v_d^{*}$, $v_q^{*}$ are
transformed back into the stationary frame using the same PLL angle
$\theta$,
\begin{equation}
    u_{r}^{ref} = v_d^{*}\cos\theta - v_q^{*}\sin\theta, \qquad
    u_{i}^{ref} = v_d^{*}\sin\theta + v_q^{*}\cos\theta,
    \label{eq:vab_star}
\end{equation}
and used as the modulation references for the PWM stage.

\section{Methodology: Case Study and Test Matrix }
\label{section3}

\begin{figure}[htbp]
\centering
\includegraphics[width=\linewidth]{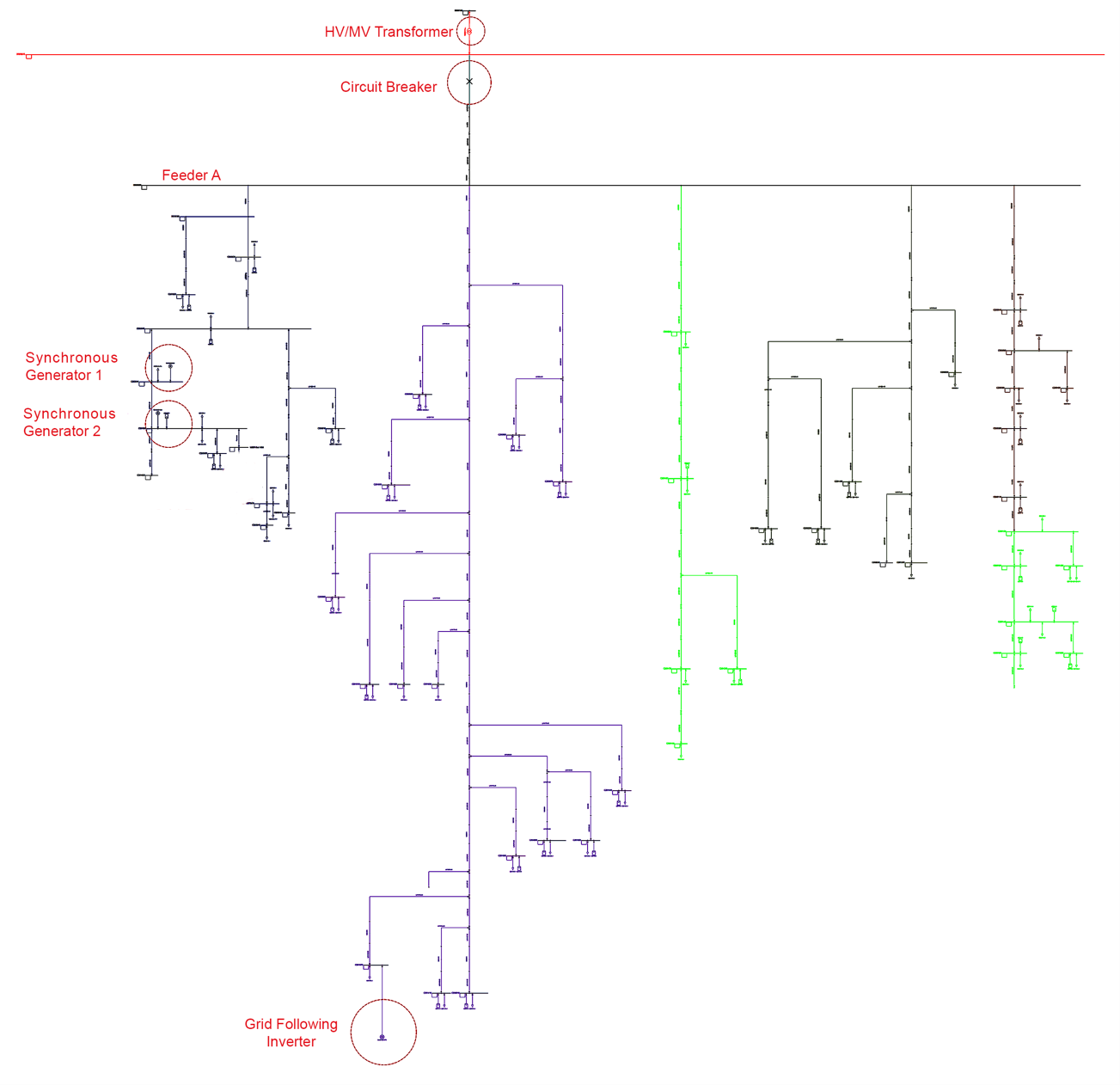}
\caption{The real MV active distribution network used as a case study.}
\label{fig:MVdistributionnetwork}
\end{figure}

The case study is based on a real 20~kV radial MV distribution network as shown in Fig-\ref{fig:MVdistributionnetwork} in Southern Italy. It hosts both traditional rotating machines and IBRs. In the original configuration, the network includes two SGs rated at 0.45~MVA each and one GFL rated at 5~MVA. Loads are modeled as constant impedances, with active and reactive absorption varying with the square of voltage magnitude \cite{b17}. They are connected to the main network through a 25~MVA HV/MV transformer.

Table \ref{tab:casedata} summarizes the network features and the transformer rated parameters. At the start of the simulation ($t=0$~s) the network operates in grid connected mode with the upstream circuit breaker closed. At $t=2$~s, the MV feeder switch opens as a DSO switching maneuver disconnecting the MV network from the main grid and supplied solely by DERs. This leads to an unintentional islanding of the isolated portion that is evaluated by voltage and frequency transients response under generation mix within the isolated portion. While future ADNs could in principle exploit DER capability to support islanded operation (e.g., during service restoration), but still such operation is presently discouraged by DSOs owing to safety hazards and equipment-damage risk. Therefore, the primary objective of this study is to investigate the conditions under which this phenomenon may occur in a real power network.

\begin{table}[t]
\caption{Case Study Network and Transformer Parameters}
\label{tab:casedata}
\centering
\footnotesize
\textbf{(a) Network Characteristics}\\[2pt]
\begin{tabular}{@{}lc@{}}
\toprule
Parameter & Value \\
\midrule
Power Flow on the Feeder [MW/Mvar] & 4.33 / 1.88 \\
Number of Loads & 54 \\
Number of Static Generators & 1 GFL \\
Number of Synchronous Generators & 2 \\
\bottomrule
\end{tabular}

\vspace{10pt}

\textbf{(b) HV/MV Transformer Electrical Parameters}\\[2pt]
\begin{tabular}{@{}lc@{}}
\toprule
Parameter & Value \\
\midrule
Rated Power [MVA] & 25 \\
Rated Voltage [kV] & 127 / 20.8 \\
Short Circuit Voltage [\%] & 14.6 \\
Copper Losses [kW] & 120 \\
No Load Current [\%] & 0.75 \\
No Load Losses [kW] & 20 \\
Vector Group & Yyn0 \\
\bottomrule
\end{tabular}
\end{table}

DER interface protections trip according to CEI~0-16 \cite{b22}, consistent with the ENTSO-E Network Code \cite{b23}: frequency range $0.95$--$1.03$~p.u.\ ($47.5$--$51.5$~Hz) and voltage range $0.85$--$1.10$~p.u.\ ($17$–-$22$~kV), respectively, based on a nominal voltage of 20 kV. If either quantity exceeds its threshold, the DER's interface protection relay detects the unintended island and opens the corresponding breaker, de-energizing the network. To represent the worst case for island persistence, DERs are set to supply exactly the power required by loads prior to disconnection, minimizing the pre-event power exchange with the upstream grid.

\subsection{Test Matrix}
To isolate the effect of DER control, both scenarios apply an identical unintentional-islanding event to the same portion of the grid. The GFL provides a common baseline, regulating active and reactive power in both scenarios. It does not participate in autonomous frequency or voltage regulation. The variable under test is the SG regulation mode: Scenario I with SGs operated at a fixed active-power setpoint and constant field excitation; Scenario II activates the SG governor and AVR, enabling active frequency and voltage regulation. Table \ref{tab:der_params} summarizes the rated power and controller parameters assigned in each scenario.

\begin{table}[t]
\caption{GFL Converter and Synchronous Generator Model Parameters}
\label{tab:der_params}
\centering
\footnotesize

\textbf{(a) GFL Power and Current Controller Parameters}\par
\begin{tabular}{lc}
\hline
Parameter & Value \\
\hline
Active power droop coefficient $K_p$ [\%]         & 5     \\
Reactive power droop coefficient $K_q$ [\%]       & 20    \\
Current filter time constant [s]                  & 0.001   \\
Gain, active current controller $K_{i,c}$ [-]      & 0.8   \\
Integrator time const., active current $T_{i,c}$ [s] & 0.001 \\
Gain, reactive current controller [-]              & 0.8   \\
Integrator time const., reactive current [s]       & 0.001 \\
\hline
\end{tabular}

\vspace{0.4em}
\textbf{(b) AVR Parameters}\par
\begin{tabular}{lc}
\hline
Parameter & Value \\
\hline
$T_r$ \; Measurement delay [s]              & 0.02  \\
$K_a$ \; Controller gain [p.u.]             & 60 \\
$T_a$ \; Controller time constant [s]       & 0.3   \\
$T_e$ \; Exciter time constant [s]          & 0.05  \\
$K_e$ \; Exciter constant [p.u.]            & 1.0   \\
$K_f$ \; Stabilization path gain [p.u.]     & 0.0   \\
$T_f$ \; Stabilization path time constant [s] & 0.5 \\
$V_{r,\text{min}}$ Controller output min. [p.u.] & $-2.15$ \\
$V_{r,\text{max}}$ Controller output max. [p.u.] & 2.15  \\
$EFD_{\text{max}}$ Exciter max. output [p.u.] & 6.0 \\
\hline
\end{tabular}

\vspace{0.4em}
\textbf{(c) Speed Governor Parameters}\par
\begin{tabular}{lc}
\hline
Parameter & Value \\
\hline
$K$ \; Actuator gain [p.u./p.u.]            & 8.0   \\
$T_1,T_2,T_3$ [s]                          & 0.1, 0.008, 0.5 \\
$T_4,T_5,T_6$ [s]                          & 0.15, 0.1, 0.12 \\
$T_D$ \; Combustion delay [s]              & 0.01  \\
Droop $D$ [p.u.]                            & 0/0.05  \\
$T_E$ \; Time const., power fdbk [s]        & 0.01  \\
$T_{\text{min}}$ \; Min. throttle [p.u.]    & $-1.0$ \\
$T_{\text{max}}$ \; Max. throttle [p.u.]    & 1.25  \\
\hline
\end{tabular}
\end{table}

\begin{table}[t]
\caption{Test-Matrix: Regulation Functions and Rated Power by Scenario}
\label{tab:testmatrix}
\centering
\footnotesize
\begin{tabular}{cp{2.9cm}p{2.9cm}}
\hline
\textbf{Sc.} & \textbf{GFL} & \textbf{SG} \\
\hline
I & Active, Reactive power control (5) & Fixed $P$ + const.\ field excitation (0.45+0.45) \\
II & Active, Reactive power control (5) & Governor + AVR control (0.45+0.45) \\
\hline
\end{tabular}
\end{table}


\section{Dynamic Performance Assessment}
\label{section4}
For each numerical simulation, voltage and frequency over the network are calculated to check their compliance with DER protection thresholds and thus identify scenarios possibly leading to undetected islanding. Table \ref{tab:testmatrix} summarizes the regulation functions and rated power in MVA assigned to each DER technology in each scenario.
\subsection{Scenario I: GFL + SGs(At Fixed P \& Constant Excitation)}
\label{scenario1}

\begin{figure}[t]
\centering
\includegraphics[width=\linewidth]{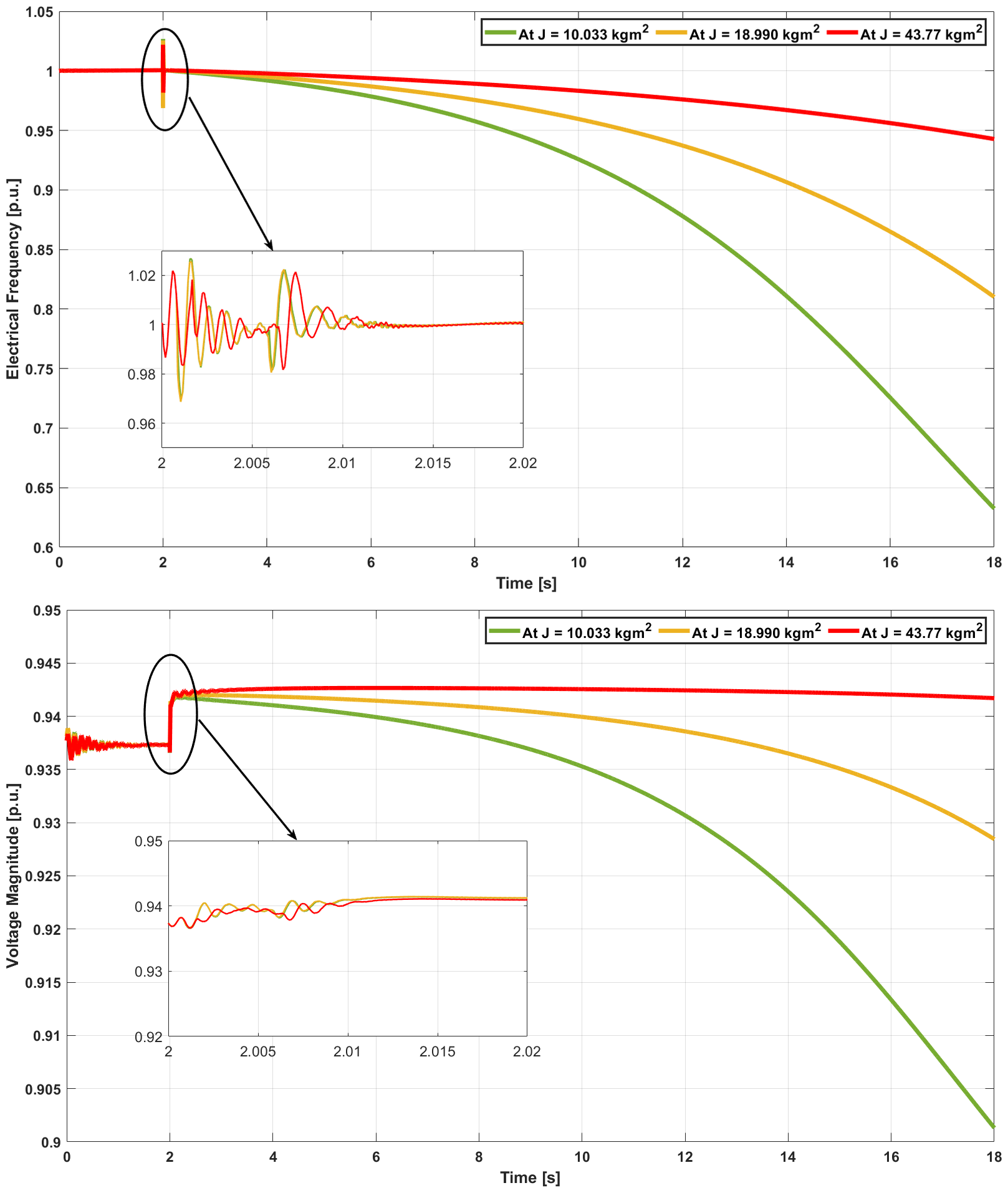}\hfill
\caption{Scenario I: frequency (upper) and voltage magnitude (lower) response for varying SG inertia $J$.}
\label{fig:s1}
\end{figure}

Following islanding at $t=2$~s, the disconnected network loses the support of the upstream grid and transitions to autonomous operation, as shown in Fig.~\ref{fig:s1}. Immediately after breaker opening, both frequency and voltage exhibit a brief oscillatory transient caused by the sudden removal of the stiff grid reference. During this period, the SG and GFL converter interact to establish a new operating equilibrium within the islanded network. Since the SG operates without governor action, the transient power imbalance following islanding is absorbed primarily by the SG rotor. Consequently, the frequency response is strongly influenced by the inertia constant $H$, with lower-inertia cases exhibiting larger and faster frequency excursions than higher-inertia cases. Owing to the short duration of the transient, the observed frequency deviations remain within acceptable limits and do not trigger protection actions. The transient is rapidly damped, settling within a few tens of milliseconds, with the damping characteristics influenced by the SG inertia. Once the transient subsides, the system response becomes dominated by the slower electromechanical dynamics of the SG.\\
The subsequent evolution of the island is primarily governed by the available inertia. Lower inertia values lead to a faster frequency decline following the active power imbalance, causing the under-frequency threshold of 0.95~p.u.\ to be reached sooner and reducing the island survival time. Conversely, higher inertia slows the rate of frequency change and prolongs operation before protection acts. Throughout all investigated cases, the voltage remains above the 0.85~p.u.\ threshold, indicating that island collapse is driven by frequency instability rather than voltage excursions. Notably, inertia values exceeding approximately $20$~kg$\cdot$m$^2$ allow the island to persist for more than 15 seconds, including the 4 second protection delay. Such durations are operationally significant from a DSO perspective, as prolonged unintentional islanding may increase the risk of protection-coordination issues and unintended out-of-synchronism auto-reclosing.



\subsection{Scenario II: GFL + SG (Performing Voltage and Frequency Control)}

\begin{figure}[t]
\centering
\includegraphics[width=\linewidth]{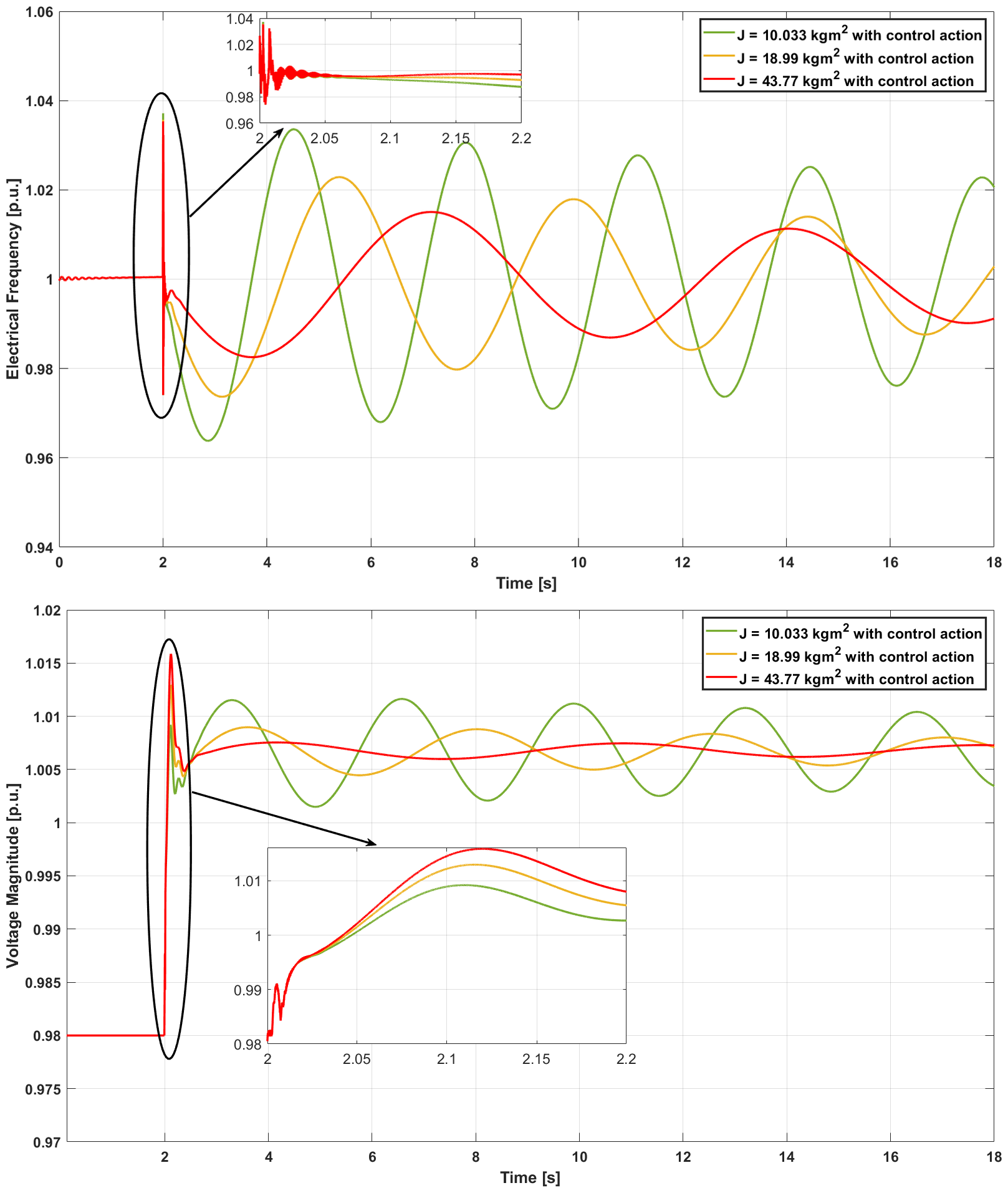}\hfill
\caption{Scenario II: frequency (upper) and voltage magnitude (lower) response for varying SG inertia $J$.}
\label{fig:s2}
\end{figure}

In this scenario, the same three representative inertia values selected from Scenario~I are used to assess the influence of SG inertia on islanding persistence when active frequency and voltage control are enabled. Following islanding at $t=2$~s, both frequency and voltage exhibit a short-duration oscillatory transient, as highlighted in the insets of Fig.~\ref{fig:s2}. This transient is caused by the sudden removal of the stiff upstream grid reference, after which the SG with $V/F$ control and GFL converter establish a new operating equilibrium within the islanded network. The short-duration frequency and voltage oscillations following islanding remain within the applicable ride-through envelopes specified the ENTSO-E RfG \cite{b23}, indicating that the transient alone does not necessitate DER disconnection. Higher inertia values provide greater damping and reduce the magnitude of the initial disturbance, while the lower-inertia case ($J=10.033$~kg$\cdot$m$^2$) exhibits the largest transient excursion.

After the initial settling period, the system transitions to a slower electromechanical response characterized by sustained oscillations whose frequency and damping depend on the SG inertia. The lowest inertia case ($J=10.033$~kg$\cdot$m$^2$) exhibits the fastest oscillatory dynamics, with a frequency overshoot of approximately 1.033~p.u. and an undershoot of 0.968~p.u., whereas higher inertia values result in slower and more damped responses. Unlike Scenario~I, all investigated inertia values within the selected range maintain both frequency and voltage within the prescribed protection thresholds, resulting in sustained unintentional islanding. This demonstrates that the SG control action is capable of balancing the islanded network and preventing the frequency and voltage excursions required for passive islanding detection.


\subsection{Results Summary and Protection Implications}
Table \ref{tab:summary} summarizes the dynamic performance indicators for both investigated scenarios, including the maximum and minimum frequency, tripping time, and islanding persistence. In Scenario I, the islanding condition is eventually cleared in all tested inertia cases; however, the duration of the islanded operation strongly depends on system inertia, ranging from 12.5 seconds at the lowest tested inertia up to 21 seconds at the highest. Although the islanding condition is not sustained indefinitely, the observed persistence time under high-inertia conditions may already be operationally significant for DSOs.

In Scenario II, where regulating synchronous generators are considered, the islanding condition persists indefinitely for all investigated inertia values. Despite variations in inertia, the frequency excursions remain limited, with maximum and minimum values within 1.037~p.u.\ and 0.974~p.u., respectively. Since both frequency and voltage are actively regulated by the synchronous generators, the system never reaches the protection thresholds, resulting in sustained unintentional islanding that conventional interface protection cannot detect.

\begin{table}[t]
\caption{Dynamic Performance Summary Across Scenarios}
\label{tab:summary}
\centering
\footnotesize
\begin{tabular}{@{}p{0.3cm}p{1.0cm}p{0.9cm}p{0.9cm}p{0.9cm}p{0.9cm}@{}}
\toprule
\textbf{Sc.} & \textbf{$J$ [kg$\cdot$m$^2$]} & \textbf{Max $f$ [p.u.]} & \textbf{Min $f$ [p.u.]} & \textbf{Trip [s]} & \textbf{Island?} \\
\midrule
\multirow{3}{*}{I} & 10.033 & 1.024 & 0.955 & 12.5 & No \\
 & 18.990 & 1.024 & 0.961 & 14.9 & No \\
 & 43.770 & 1.029 & 0.983 & 21.0 & No \\
\multirow{3}{*}{II} & 10.033 & 1.037 & 0.974 & -- & Yes \\
 & 18.990 & 1.035 & 0.974 & -- & Yes \\
 & 43.770 & 1.034 & 0.974 & -- & Yes \\
\bottomrule
\end{tabular}
\end{table}

These findings demonstrate that the presence of active frequency and voltage regulation within the islanded network can effectively suppress the frequency and voltage deviations typically relied upon by conventional passive islanding protection schemes. Since SGs achieve this regulation through comparatively slow electromechanical processes, with governor and AVR dynamics acting on timescales of several seconds, their ability to sustain the island while maintaining operating conditions within protection thresholds is particularly significant. Consequently, converter-based DERs equipped with grid-forming control, which can regulate voltage and frequency on much faster sub-cycle to sub-second timescales, are expected to exhibit similar or even stronger island-sustaining capabilities. This insight highlights the need for a dedicated assessment of GFM-based resources

\section{Conclusion and Future Work}
\label{section5}
This paper has analyzed the dynamic behavior of a GFL converter and SGs operated with and without active frequency/voltage regulation, during the transition from grid-connected to islanded operation in a real MV distribution network. Detailed EMT models of both technologies were developed and implemented in DIgSILENT PowerFactory and validated against a real network. The numerical simulations show that DERs operating at a fixed power setpoint pose a limited risk of sustained unintentional islanding, since the isolated network drifts out of the protection thresholds within tens of seconds depending on system inertia. In contrast, enabling active frequency and voltage regulation on the SGs, even with the comparatively slow response of electromechanical governor and AVR control is sufficient to sustain the island indefinitely without triggering conventional protection regardless of inertia.

These findings indicate that the critical factor governing unintentional islanding persistence is the presence of active regulation on the islanded portion, rather than the specific rotating or power-electronic technology providing it. Since grid-forming converters are designed to regulate frequency and voltage with a much faster and more precise response than electromechanical SG control, they are expected to pose an equal or greater islanding risk than the regulating-SG case analyzed here. Future work will therefore extend this real-network validated benchmark to grid-forming converters, quantifying their islanding persistence and detection-time implications relative to the SG and GFL baselines established in this paper. Additional studies will investigate the sensitivity of these results to network conditions, including load composition, DER penetration, and topology, to further quantify the robustness and generality of the observed islanding behavior. Finally, hardware-in-the-loop validation and techno-economic assessment of grid-forming inverter deployment for intentional islanding and black-start support will be pursued at the distribution level.

\vspace{12pt}

\end{document}